# Simultaneous anti-bunched and super-bunched photons from a GaAs Quantum dot in a dielectric metasurface

**Sanghyeok Park,[1,2] Oleg Mitrofanov,[3] Kusal M. Abeywickrama,[4] Samuel Prescott,[3] Jaeyeon Yu,[1,2] Stephanie C Malek,[1,2] Hyunseung Jung,[1,2] Emma Renteria,[5] Sadhvikas Addamane,[1,2] Alisa Javadi,[4] Igal Brener,[1,2,*] and Prasad P Iyer[1,2,*]**

[1]*Sandia National Laboratories, Albuquerque, New Mexico 87185, United States*
[2]*Center for Integrated Nanotechnologies, Sandia National Lab, Albuquerque, New Mexico 87185, United States*
[3]*University College London, Electronic and Electrical Engineering, London WC1E 7JE, U.K*
[4]*Department of Electrical and Computer Engineering, University of Oklahoma, Norman, Oklahoma 73019, United States*
[5]*Center for High Technology Materials, University of New Mexico, Albuquerque, New Mexico 87185, United States*
[*ibrener@sandia.gov](mailto:ibrener@sandia.gov), [ppadma@sandia.gov](mailto:ppadma@sandia.gov)

Semiconductor quantum dots host a rich manifold of excitonic complexes, including neutral excitons that emit anti-bunched single photons and charged exciton complexes capable of producing super-bunched photons via cascade emission. Accessing both emission regimes from a single emitter would open routes to novel quantum protocols, including advanced quantum imaging. In practice, however, emission from charged exciton complexes is intrinsically weak, often orders of magnitude dimmer than neutral excitons, placing simultaneous dual-mode operation out of reach. Here, we overcome this limitation by embedding the quantum dot in a dielectric Mie-resonant metasurface that provides order-of-magnitude photoluminescence enhancement across both neutral and charged exciton transitions of a single GaAs quantum dot. Under identical non-resonant pumping conditions, the emission from the neutral exciton yields anti-bunched emission ($g^{(2)}(0) < 0.5$) and the emission from positively charged exciton complexes shows super-bunched emission ($g^{(2)}(0) > 3.5$) with comparable count rates (~12 kHz). Crucially, super-bunching emerges only when charged exciton emission spectrally overlaps with the Mie resonances and vanishes in unpatterned slabs, demonstrating that photonic engineering, is essential for accessing these weak quantum light states. These results demonstrate a scalable, position-tolerant platform for harnessing the full excitonic structure of solid-state emitters.

**Introduction**

Quantum information technologies such as quantum communication, sensing, and computing have advanced rapidly with the development of sophisticated quantum light sources. Among these emerging technologies, quantum imaging employing non-classical light, such as single photons, entangled photons, and squeezed photons, has demonstrated the ability to surpass classical limits of resolution, sensitivity, and information extraction [1-5]. Techniques such as ghost imaging [6], remote sensing [7], and sub-shot-noise imaging [8] have demonstrated the ability to reconstruct images with fewer photons, reduced noise, and improved contrast by exploiting tailored photon statistics. Recently, it has been suggested that the simultaneous use of anti-bunched and super-bunched photons within a single protocol could unlock new regimes of quantum imaging, such as polarization sensitive- and edge-enhanced imaging [9]. More broadly, quantum light sources capable of producing distinct photon statistics under identical pumping conditions could provide a flexible platform for exploring alternative strategies in quantum communication, sensing, and information processing.

Epitaxially grown semiconductor quantum dots (QDs) are bright, scalable solid-state quantum light sources which can host multiple excitonic states with distinct quantum-optical signatures [10–13]. Isolated excitons (neutral $X^0$ or charged $X^+/X^-$) emit anti-bunched photons [14, 15], while biexcitons ($XX^0$, $XX^+/XX^-$) may produce super-bunched photons via cascade emission processes [16, 17]. This intrinsic diversity positions QDs as ideal candidates for versatile quantum light sources. However, realizing simultaneous access to both emission regimes remain elusive. In a QD without applied gate bias, charged exciton complexes typically exhibit weak photoluminescence (PL) intensities, often a few orders of magnitude lower than neutral excitons [18, 19]. The challenge is enhancing weak charged exciton emission without compromising neutral exciton performance, while accommodating for the inherent spectral and spatial inhomogeneities of epitaxially grown QDs.

Selective enhancement or suppression of excitonic emission and efficient out-coupling into free space can be realized via photonic environment engineering [20-27]. Dielectric metasurfaces supporting relatively broad Mie resonances provide a particularly scalable solution. Unlike high-Q cavities that demand precise spectral tuning and nanometer-scale emitter positioning, Mie resonances offer moderate Q-factors with tens of nanometers bandwidths, allowing simultaneous enhancement of multiple spectrally separated transitions [26, 28-30]. Their uniform spatial mode profiles make the enhancement robust to QD position and size variations, eliminating the need for post-growth selection or tuning [31, 32]. Metasurfaces have already been used to boost single-photon emission from epitaxial and colloidal QDs and to provide spatially and spectrally stable emission control [19, 29, 33]. However, most demonstrations have targeted a single excitonic species. By leveraging their bandwidth and robustness, Mie metasurfaces can enhance emission from the manifold of exciton complexes simultaneously, independent of the position and size of QDs, enabling a new approach that harnesses the full excitonic structure for multifunctional quantum light generation.

Here, we demonstrate simultaneous generation of anti-bunched and super-bunched light from a single local droplet etched GaAs QD using such a metasurface platform, which provides an order-of-magnitude PL enhancement across both neutral excitons and charged exciton complexes, enabling access to emission regimes that are typically too weak to utilize. Through magneto-photoluminescence spectroscopy, we identify neutral excitons ($X^0$) and positively charged complexes as the origin of anti-bunched and super-bunched emission, respectively. Photon correlation measurements reveal that super-bunching signatures emerge only when charged exciton emission spectrally overlaps with the electric- and magnetic-dipole resonances of the metasurface, confirming the essential role of photonic engineering in accessing these weak transitions. Under identical non-resonant pumping conditions, we achieve comparable photon count rates (~10-12 kHz) for both anti-bunched and super-bunched emission, demonstrating the practical viability of this dual-mode quantum light source. These results establish a scalable, position-independent approach to harnessing the full excitonic manifold of semiconductor QDs.

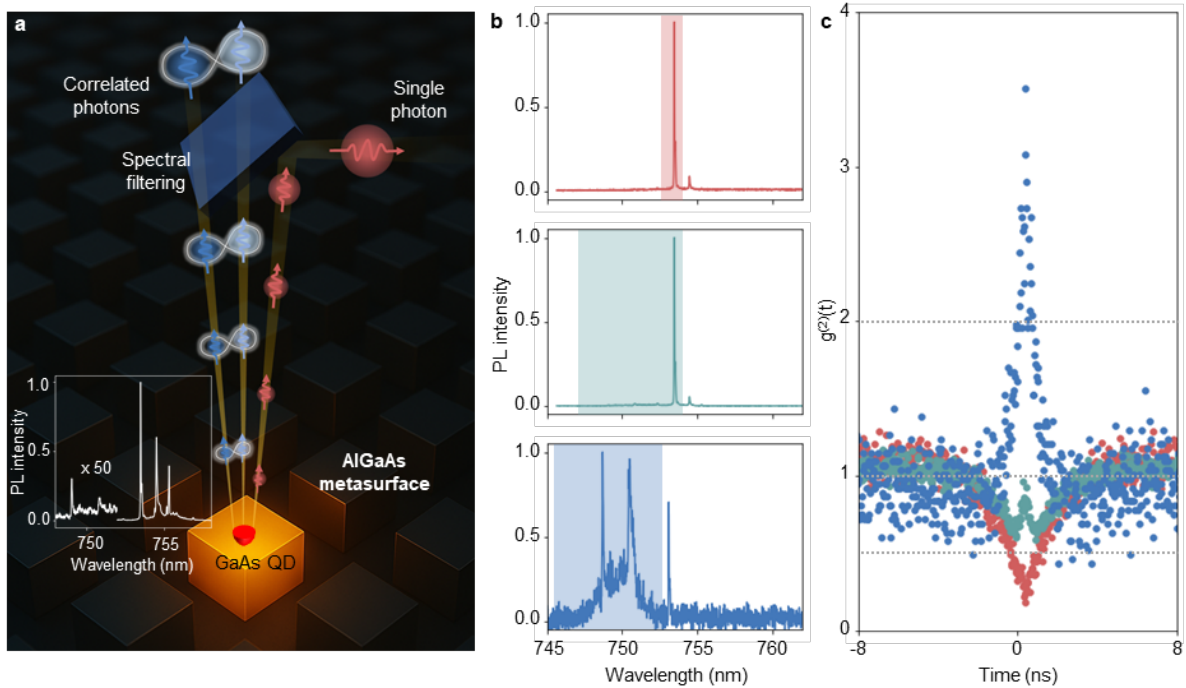

Figure 1 | **Simultaneous generation of anti-bunched- and super-bunched photon states from a GaAs quantum dot in a metasurface**. (a) Illustration of a GaAs quantum dot (QD) embedded in an AlGaAs metasurface. The structured photonic environment enables emission of both single photons (red arrow) and time-correlated photons (super-bunched) states (blue arrows). *Inset:* Normalized photoluminescence (PL) spectrum of a QD under non-resonant excitation. (b) Spectrally filtered and normalized PL spectra from a single QD. Shaded regions indicate the selected spectral windows. The top spectrum (red) isolates emission from the dominant peak at 753 nm; the middle spectrum (green) includes both the dominant peak and weaker short-wavelength features; the bottom spectrum (blue) isolates the short-wavelength emission near 750 nm. (c) Corresponding Hanbury Brown and Twiss second-order correlation measurement results demonstrating single-photon emission, $g^{(2)}(0) < 0.5$, when we isolate the brightest exciton peak (red curve) using an optical filter, and emission from exciton complexes, $g^{(2)}(0) > 2$ when we isolate shorter wavelength emission (blue curve). Emission containing both features in the spectrum (green curve) shows a peak ($g^{(2)}(0) \sim 1$) inside an anti-bunching dip ($g^{(2)}(0) < 1$).

**Results**

**Second-order correlation measurements of emission from exciton complexes:** Figure 1a is an artistic illustration of the core concept: a GaAs QD embedded within an AlGaAs-based Mie resonant metasurface. This structured photonic environment simultaneously enhances the emission of both single photons depicted with the red arrows and time-correlated multiphoton (super-bunched) states depicted with the blue arrows from the same QD. The Mie resonators are $Al_{0.4}Ga_{0.6}As$ cuboids arranged in a square lattice on a sapphire substrate fabricated via a flip-chip process [34]. Details of the fabrication process are outlined in the Methods section. The metasurface was engineered to support electrical dipole (ED) and magnetic dipole (MD) modes at approximately the wavelength of QD emission (750 nm, Fig. 1a, Inset) by varying the width of cuboids and the period of the square lattice. By aligning the metasurface modes with the QD emission spectrum, we enhance the PL from the QD by an order of magnitude across the emission wavelength range of exciton complexes (Supplementary Information S1) [28, 29].

The emission from an isolated GaAs QD in the metasurface exhibits both anti-bunched and super-bunched photon statistics depending on applied spectral filtering. This dual functionality arises from the metasurface's simultaneous enhancement of PL emission from multiple exciton complexes. Figure 1b shows PL spectra of the QD under continuous-wave non-resonant excitation (520 nm) after spectral filtering has been applied using two band-edge filters (Methods), while Fig. 1c displays the corresponding second-order photon correlation $g^{(2)}(\tau)$ measurements using a Hanbury Brown and Twiss (HBT) setup. When the emission from the brightest peak is spectrally isolated, a pronounced anti-bunching behavior is observed ($g^{(2)}(0)<0.5$). In contrast, the emission at shorter wavelengths results in super-bunching behavior ($g^{(2)}(0)>2$). We observe a clear transition from the anti-bunched state to super-bunched state. Both distinct photon statistics showed similar levels of photon count rates (~ 12 kHz) at the same pump power density (18 W/cm$^2$). We therefore can selectively access different types of quantum light from an isolated QD in the metasurface via the application of spectral filtering.

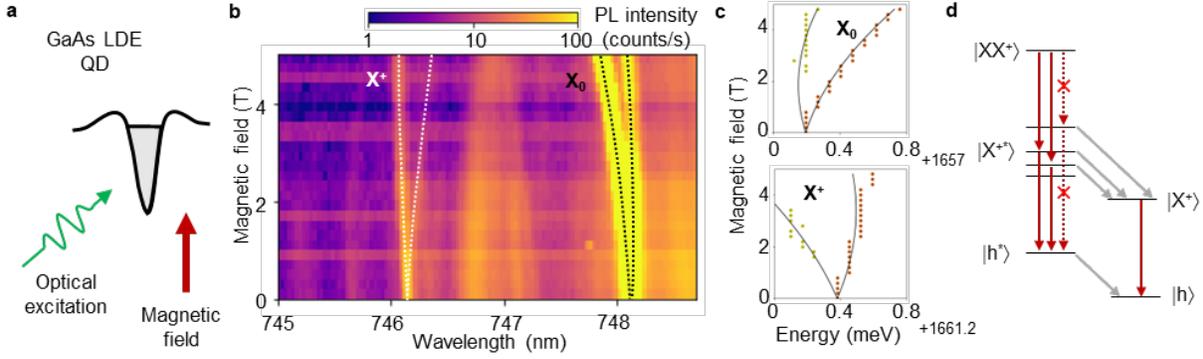

Figure 2 | **Magneto-photoluminescence characterization of excitonic states in GaAs QDs within metasurfaces**. (a) Schematic of the experimental setup: a QD is excited by a 532 nm non-resonant laser (green arrow) under an external magnetic field (red arrow) applied in the Faraday geometry. (b) Magneto-PL map showing QD emission as a function of wavelength (horizontal axis) and magnetic field (vertical axis, up to 5 T). Two excitonic transitions are identified: the neutral exciton ($X^0$) near 748 nm and the positively charged exciton ($X^+$) near 746 nm. Dashed lines correspond to fits based on the combined Zeeman effect and diamagnetic shift model. (c) Extracted energy positions of $X^0$ (top) and $X^+$ (bottom) states as a function of magnetic field, fitted with the model $E(B) = E_0 \pm \frac{1}{2} g \mu_B B + \gamma B^2$. (d) Possible cascade recombination processes within the QD, where the positively charged exciton $X^+$ arises from charged biexciton states, enabling cascaded decay channels (red arrows) The solid arrows indicate radiative decay channels and the dashed arrows indicate forbidden decay channels. The gray arrows indicate non-radiative transitions.

**Magneto-photoluminescence from GaAs Quantum Dots**: We conducted magneto-photoluminescence (magneto-PL) measurements to confirm that the origin of the observed super-bunching is positively charged exciton complexes. These experiments were carried out at 4 K with an external magnetic field applied between 0 and 5 T (Tesla) in the Faraday geometry and non-resonant optical excitation (Figure 2a). The details are in Methods. The setup enables resolving the energy splitting and shifts of excitonic transitions as a function of the magnetic field strength, thus revealing the nature and charge state of the QD emission [34, 35]. The magneto-PL spectra exhibit the key excitonic emission lines: the charge-neutral exciton at wavelength ∼ 748 nm (1658 meV), which shows anti-bunched emission, and the positively charged exciton ($X^+$) at wavelength ∼ 746 nm (1662 meV), which shows super-bunched emission. Figure 2b shows the emission spectrum of the QD as a function of magnetic field (vertical axis) and wavelength (horizontal axis). By fitting of the magnetic field-dependent energy shifts, we extracted both the Zeeman splitting and diamagnetic shifts following the established model for QDs as $E(B) = E_0 \pm \frac{1}{2}g\mu_B B + \gamma B^2$, where g is the Landé g-factor and γ is the diamagnetic coefficient. Two distinct excitonic features are marked: the charge-neutral exciton $X^0$ and the positively charged exciton $X^+$. The dashed curves overlaying the map correspond to fits based on the Zeeman effect and diamagnetic shift models, allowing extraction of g-factors and diamagnetic coefficients. Figure 2c displays the extracted energies of the $X^0$ and $X^+$ states as a function of magnetic field. The fitted value of g (1.67) and γ (13.1 µeV/T²) confirm that this state exhibits typical behavior for the $X^0$ state in GaAs QDs [34]. We conclude that the shorter wavelength peak (~ 746 nm) originates from the positively charged exciton state owing to the negative binding energy [36] and the anomalous diamagnetic shift [34, 37]. For small QDs, the hole-hole repulsion can dominate the energy renormalization in positively charged exciton, pushing the binding energy to negative values [36], while the negative sign of the diamagnetic coefficient γ (-10.0 µeV/T²), known as the anomalous diamagnetic shift, has been associated with the positively charged exciton in GaAs QD [34].

Figure 2d illustrates possible cascade recombination processes within the QD. Unlike the neutral biexciton cascade emission, the charged biexciton cascade emission includes decay channels with the excited charge carrier in the p-shell (excited charged exciton, $|X^{+*}\rangle$). The charged biexciton has two holes and electrons in the s-shell and an additional hole in the p-shell. At first, one electron-hole pair (s-shell) decays and the excited charged exciton $|X^{+*}\rangle$ remains. It can have four different spin configurations depicted as four levels in Fig 2d [16, 17]. Due to the spin selection rules, two radiative decay channels (from $|XX^+\rangle$ to singlet $|X^{+*}\rangle$ state and to triplet $|X^{+*}\rangle$ state with $m_z$ = 3/2) are forbidden. The cascaded decay consisting of the radiative decay from $|XX^+\rangle$ to $|X^{+*}\rangle$ and from $|X^{+*}\rangle$ to $|h^*\rangle$ can generate the time-correlated photon pairs. [16, 17]. Unfortunately, the spectral resolution is not enough to fully identify the origins of emission lines with clearly resolving the Zeeman splitting and diamagnetic shifts of the peaks around 747 nm. But the quadratic pump power dependence led us to conclude that at least one PL peak comes from the charged biexciton (see Supplementary Information S2). The cascade emission process can produce highly correlated photon pairs, consistent with the observed super-bunching (Supplementary Information S3).

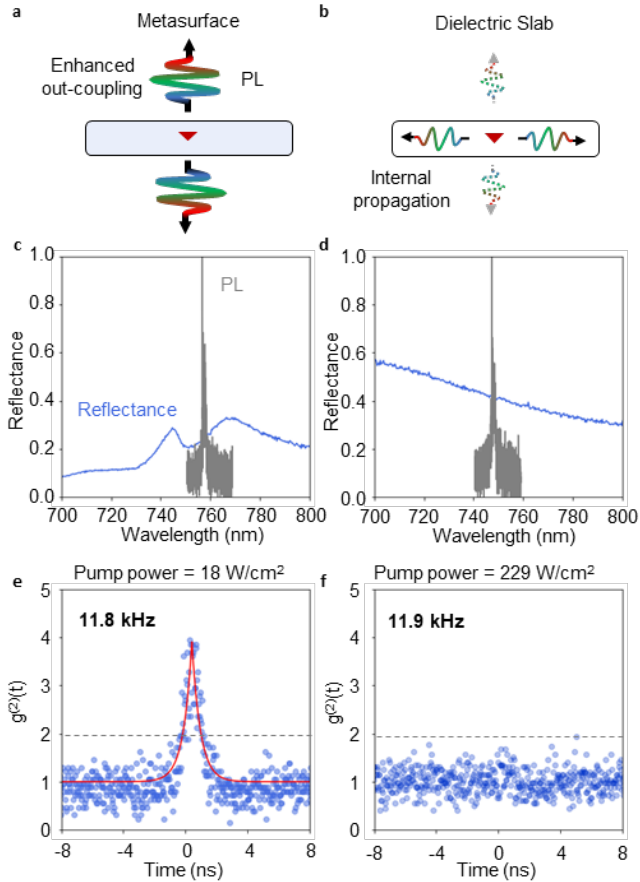

Figure 3 | **Role of photonic environment in enabling super-bunched emission from charged exciton complexes.** (a, b) Schematic of PL emission from (a) a QD in the metasurface and (b) a QD in the dielectric slab (no photonic structure). The metasurface can enhance the out-coupling efficiency of PL into the free-space and radiative recombination rate. (c) PL spectrum from the positively charged exciton complexes under a pump power density of 18 W/cm² and reflectance of the Mie metasurface. (d) PL spectrum under a pump power density of 229 W/cm² and reflectance of the 140-nm-thick $Al_{0.4}Ga_{0.6}As$ slab. (e) Second-order correlation measurement of the PL from the QD embedded in Huygens' metasurface. Clear super-bunching behavior with a peak $g^{(2)}(0) = 3.90$ is shown. (f) Second-order correlation measurement of the PL from the QD embedded in dielectric slab. The emitted photon shows $g^{(2)}(0) < 2$. A ten times higher excitation power (229 W/cm²) is required to reach a similar photon count rate (11.9 kHz) with the metasurface case (11.8 kHz).

**Enhanced out-coupling of emission from charged excitons in GaAs Quantum dots:** Super-bunched photon emission from positively charged exciton complexes is observed only when the GaAs QD is embedded in the metasurface, demonstrating that engineered photonic environments play an essential role in accessing this non-classical emission regime. To understand this dependence, we compare GaAs QDs in two distinct configurations: embedded in a Huygens' metasurface and an unpatterned $Al_{0.4}Ga_{0.6}As$ slab of identical thickness (Fig. 3a,b). The metasurface supports two overlapping Mie resonances, an electric dipole (ED) mode and a magnetic dipole (MD) mode, whose sub-wavelength periodicity directs emission preferentially in the upward and downward directions [28, 29, 38]. Figure 3(c) shows the reflectance spectrum exhibiting these resonant features overlapping with the QD emission wavelengths. In contrast, the unpatterned slab displays a featureless reflectance spectrum (Fig. 3d). Due to the large refractive index contrast between AlGaAs (n ~ 3.5) and air, most emission from QDs in the slab is guided laterally via total internal reflection, with minimal out-coupling to free space [28, 29]. These stark differences in photonic environments, enhanced out-coupling in the metasurface versus the guided emission with negligible free-space extraction in the slab, have profound consequences for the observable photon statistics.

The photonic environment provides a crucial role to access the super-bunching phenomenon. We measured second-order photon correlation functions $g^{(2)}(\tau)$ from charged exciton complexes in both environments, filtering spectrally to isolate emission from positively charged exciton complexes. Figures 3(e) and (f) show the second-order photon correlation function for the emission from the positively charged QD exciton complexes in each photonic environment. While both exhibit comparable total photon count rates (~12 kHz), the excitation conditions differ significantly: the unpatterned slab requires a 10 times higher excitation power density (229 W/cm²) compared to the metasurface (18 W/cm²). The metasurface-embedded QD exhibits strong super-bunching with $g^{(2)}(0) = 3.90$, indicating temporally correlated photon bursts characteristic of cascade emission from charged biexciton complexes. In stark contrast, the slab-embedded QD shows $g^{(2)}(0) < 2$, remaining within the classical regime with no appreciable bunching. The drastic pump condition difference is important to understand why super-bunching is observed only in the metasurface case. At the high pump powers needed to detect emission from the slab with same count rates, spectral broadening, thermal effects, and activation of uncorrelated background states dilute temporal correlations, washing out the super-bunching signature. The metasurface's effective photon outcoupling into free-space allows access to the intrinsic cascade dynamics at low pump powers where super-bunching is preserved. The spectral alignment between the QD emission, particularly from weak charged exciton transitions, and the overlapping ED and MD resonances is essential for this effect. We confirmed this behavior in additional metasurface designs with different Mie resonance configurations, all showing super-bunched emission from charged exciton complexes (Supplementary Information S4). Beyond brightness enhancement via broadband low-Q Mie resonances, metasurfaces can also provide high-Q resonant filtering for spectral and polarization-selective isolation of specific transitions [39, 40] This dual capability, simultaneous enhancement of multiple transitions combined with selective filtering, establishes metasurface-embedded QDs as a versatile platform for engineering and routing diverse quantum light states on demand [10, 12].

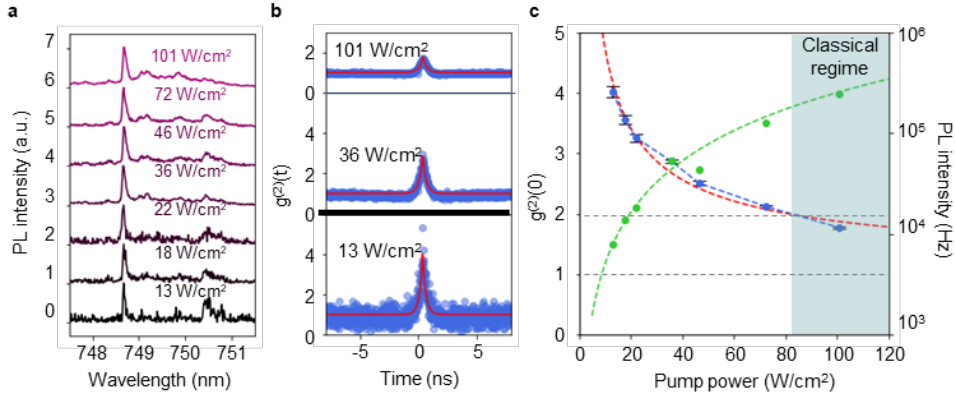

Figure 4 | **Power dependence of PL from the GaAs QD in the Mie metasurface** (a) Power-dependent PL spectra recorded from 13 to 101 W/cm². (b) Second-order correlation measurements at selected power densities (13, 36, and 101 W/cm²). (c) Extracted $g^{(2)}(0)$ values as a function of pump power P. It follows a sublinear inverse trend, approximately modeled as $g^{(2)}(0) \sim 1+a/P^{0.63}$.

**Pump power dependence of super-bunched emission:** Beyond enabling observation of super-bunched emission, the metasurface's low operating power regime is critical for preserving strong quantum correlations. To quantify this effect, we measured the power dependence of super-bunching from the positively charged exciton complexes. Figure 4(a) shows the PL spectra from the positively charged exciton complexes of the QD in the metasurface under increasing continuous-wave non-resonant pump power ranging from 13 W/cm² to 101 W/cm². The corresponding second-order correlation functions $g^{(2)}(\tau)$ at selected pump power densities (13, 36, and 101 W/cm²) are shown in Figure 4(b). At low pump power (13 W/cm²), the correlation function exhibits a pronounced peak at zero delay time with $g^{(2)}(0)$ exceeding 4, indicating strong super-bunching and highly correlated multiphoton states from the cascade emission. As pump power increases, this peak systematically diminishes, and the correlation function gradually approaches the classical regime ($g^{(2)}(0) \leq 2$).

Two competing physical mechanisms account for the observed reduction in photon correlations at higher pump powers. First, spectral broadening of excitonic lines occurs due to increased thermal fluctuations and charge accumulation [41-44], which reduces temporal coherence of the cascade emission. Second, higher excitation activates otherwise weak excitonic states that contribute to uncorrelated photon background, decreasing the relative contribution of correlated photon pairs from the cascade decay process. Together, these effects systematically dilute the temporal correlations, resulting in the observed decrease of $g^{(2)}(0)$. Importantly, this behavior reinforces our earlier observation from the unpatterned slab: there, achieving detectable emission required pump powers more than an order of magnitude higher, at which point these destructive effects completely suppress super-bunching. To quantify the degradation of photon correlations with increasing pump intensity, we fit the experimental data to a phenomenological model: $g^{(2)}(0) = 1 + a/P^b$, where a and b are empirical fit parameters. Figure 4(c) shows the evolution of $g^{(2)}(0)$ as a function of pump power, clearly following a sublinear decay with b ≈ 0.63 (a ≈ 16 (W/cm$^2$)$^{0.63}$). A rigorous theoretical explanation for this specific exponent would require solving complex rate equations accounting for many-body interactions, charge capture dynamics, and carrier diffusion [45-47], an analysis beyond the scope of this work. Importantly, the metasurface's resonant enhancement enables access to the intrinsic super-bunching behavior at sufficiently low pump powers where quantum correlations remain strong. At higher powers within the operating range, we still achieved photon count rates exceeding 100 kHz while maintaining $g^{(2)}(0) > 2$, demonstrating the practical utility of this platform for generating non-classical light from weak charged exciton states.

**Conclusion**

We have demonstrated simultaneous generation of anti-bunched and super-bunched photon states from a single GaAs QD under identical non-resonant pumping conditions, enabled by an AlGaAs-based dielectric metasurface platform. Using magneto-PL spectroscopy, we identified neutral excitons ($X^0$) as the source of anti-bunched emission and positively charged biexciton complexes ($XX^+$) as the origin of super-bunched photons with $g^{(2)}(0)$ exceeding 3.5. The metasurface provides an order-of-magnitude PL enhancement across both neutral and charged excitonic transitions via broadband Mie resonances, enabling comparable photon count rates (~12 kHz) for both emission regimes at low pump powers. Crucially, photon correlation measurements revealed that super-bunching emerges only when charged exciton emission spectrally overlaps with the metasurface resonances, confirming the essential role of photonic engineering in accessing these intrinsically weak transitions. Power-dependent measurements demonstrate that low operating powers enabled by resonant enhancement preserve strong quantum correlations, as higher excitation dilutes temporal correlations through spectral broadening and uncorrelated background emission. The metasurface approach overcomes fundamental limitations of conventional photonic platforms by achieving broadband enhancement without requiring precise spectral tuning or deterministic emitter positioning, establishing a scalable route to harnessing the full excitonic manifold of semiconductor QDs and advanced single photon emitters [48]. This work establishes photonic metasurfaces as a versatile platform for engineering quantum light states on demand, opening pathways for advanced quantum information processing and photonic technologies that fully exploit the rich excitonic structure of solid-state emitters. [9].

**Methods:**

*Reflectance spectroscopy:* Reflectance spectra from both the metasurface and the unpatterned dielectric slab were acquired using a broadband tungsten-halogen white light source and a visible-range spectrometer (Ocean Optics, Jazz). The incident beam was focused onto the sample using a 50 mm focal length plano-convex lens. For normalization, reflectance from the metasurface was referenced against that of a silver mirror. Prior to normalization, dark counts were subtracted from both the metasurface and mirror spectra.

*Photoluminescence spectroscopy:* PL measurements were conducted with the sample cooled to 10 K using a closed-cycle cryostat (Montana Instruments). Non-resonant excitation was provided by a 520 nm diode laser, focused to a ~1 μm spot using a microscope objective (NA = 0.4). The emitted PL signal was collected through the same objective and analyzed with a 50 cm focal length spectrometer equipped with an 1800 groove/mm grating and a 1340 × 100 back-illuminated CCD array detector (Teledyne Princeton Instruments). To filter the PL from exciton complexes in the spectral domain, a pair of long pass and short pass filters was employed (TLP01-790, TSP01-790, AVR Optics). To measure the second-order correlation function, photon arrival times were recorded using a pair of superconducting nanowire single-photon detectors (Quantum Opus) and a time-correlated single-photon counting system (PicoHarp 300, PicoQuant). Time-tagged photon correlation measurements were performed using pyscan, an open-source measurement suite developed by the Center for Integrated Nanotechnologies (available at github.com/sandialabs/pyscan).

*MagnetoPL spectroscopy:* Magneto PL measurements were conducted with the sample cooled to 4 K using a He cryostat (AttoDRY800, Attocube). Non-resonant excitation was provided by a 532 nm diode laser, using an aspheric lens (f = 3.1mm, NA = 0.7, C330TMD-B, Thorlabs). The emitted PL signal was collected through the same lens and analyzed with a spectrometer equipped with a 1200 groove/mm grating and a CCD array detector (Horiba Scientific Instruments). To measure the second-order correlation function, photon arrival times were recorded using a pair of superconducting nanowire single-photon detectors (Single Quantum) and a time tagger system (Swabian Instruments). To filter the PL from exciton complexes in the spectral domain, a pair of long pass and short pass filters was employed (TLP01-790, TSP01-790, AVR Optics)

*Numerical Calculation*: The reflectance spectra were calculated with the rigorous coupled-wave analysis (open source python package: grcwa [49]) and the photonic band structures were calculated with the guided-mode expansion (open source python package: legume [50]). The employed refractive index of $Al_{0.4}Ga_{0.6}As$ was from [51].

*Fabrication:* Samples were grown by molecular beam epitaxy on semi-insulating GaAs (100) substrates (VA1166). After in situ oxide desorption at 630 °C, growth was initiated at 600 °C. The heterostructure consisted of a 300 nm GaAs buffer, a 500 nm $Al_{0.75}Ga_{0.25}As$ etch-stop layer, and a 140 nm $Al_{0.4}Ga_{0.6}As$ layer with local droplet etched GaAs quantum dots (LDE GaAs QDs) at its center, which are encapsulated by 5 nm GaAs layers.

QD formation was achieved by aluminum droplet etching followed by migration-enhanced epitaxy. Growth was interrupted at the midpoint of the $Al_{0.4}Ga_{0.6}As$ layer, and the substrate temperature was raised to 620 °C under an arsenic soak. After removing excess arsenic, 0.6 monolayers of Al were deposited and annealed under low-arsenic conditions to form nanovoids. GaAs QDs were subsequently grown by alternating Ga and As deposition over 24 cycles, followed by annealing.

Metasurfaces were patterned using electron-beam lithography on dual-layer PMMA resist (PMMA 495/950 A2) and reactive ion etching in $BCl_3/Cl_2/Ar/N_2$ plasma (10, 10, 10, and 3.5 cm$^3$/min, respectively). After resist removal, the samples were flip-chip bonded onto sapphire substrates using epoxy (EPO-TEK 353ND). The GaAs substrate was removed by mechanical lapping and selective wet etching with $NH_4OH/H_2O_2$, followed by removal of the $Al_{0.75}Ga_{0.25}As$ etch-stop layer using HCl, leaving a QD metasurface layer on the sapphire substrate. Details are in [29].

**Acknowledgement**

This work was supported by the U.S. Department of Energy (DOE), Office of Basic Energy Sciences, Division of Materials Sciences and Engineering, and performed, in part, at the Center for Integrated Nanotechnologies, an Office of Science User Facility operated for the U.S. DOE Office of Science. S.P. was supported by the EPSRC (EP/S022139/1). A.J. And K.A. acknowledge funding from the Office of Basic Energy Sciences through QuPIDC Energy Frontier Research Center under Award No. DE-SC0025620 Sandia National Laboratories is a multi-mission laboratory managed and operated by National Technology and Engineering Solutions of Sandia, LLC., a wholly owned subsidiary of Honeywell International, Inc., for the U.S. DOE's National Nuclear Security Administration under contract DE-NA0003525. This paper describes objective technical results and analysis. Any subjective views or opinions that might be expressed in the paper do not necessarily represent the views of the U.S. DOE or the United States Government. We thank Dr. Ashwin Kumar Boddeti for his valuable insights and helpful discussions.

# References

(1) Seitz, P.; Theuwissen, A. J. P. Single-Photon imaging, Springer. 2011.

(2) Moreau, P.-A; Toninelli, E.; Gregory, T.; Padgett, M. J. Imaging with quantum states of light. Nature Reviews Physics 2019, 1 (6), 367−380.

(3) Defienne, H.; Bowen, W. P.; Chekhova, M.; Lemos, G. B.; Oron, D.; Ramelow, S.; Treps, N.; Faccio, D. Advances in quantum imaging. Nature Photonics 2024, 18 (10), 1024−1036.

(4) Gatto Monticone, D.; Katamadze, K.; Traina, P.; Moreva, E.; Forneris, J.; Ruo-Berchera, I.; Olivero, P.; Degiovanni, I. P.; Brida, G.; Genovese, M. Beating the Abbe Diffraction Limit in Confocal Microscopy via Nonclassical Photon Statistics. Physical Review Letters 2024, 113 (14), 143602.

(5) Tenne, R.; Rossman, U.; Rephael, B.; Israel, Y.; Krupinski-Ptaszek, A.; Lapkiewicz, R.; Silberberg, Y.; Oron, D. Super-resolution enhancement by quantum image scanning microscopy. Nature Photonics 2019, 13 (2), 116−122.

(6) Pittman, T. B.; Shih, Y. H.; Strekalov, D. V.; Sergienko, A. V. Optical imaging by means of two-photon quantum entanglement. Physical Review A 1995, 52 (5), 3429−3432.

(7) Lemos, G. B.; Borish, V.; Cole, G. D.; Ramelow, S.; Lapkiewicz, R.; Zeilinger, A. Quantum imaging with undetected photons. Nature 2014, 512 (7515), 409−412.

(8) Brida, G.; Genovese, M.; Ruo-Berchera, I. Experimental realization of sub-shot-noise quantum imaging. Nature Photonics 2010, 4 (4), 227−230.

(9) Ye, Z.; Wang, H. -B.; Xiong, J.; Wang, K. Antibunching and superbunching photon correlations in pseudo-natural light, Photonics Research 2022, 10 (3), 668-677.

(10) Heindel, T.; Kim, J.-H.; Gregersen, N.; Rastelli, A.; Reitzenstein, S. Quantum dots for photonic quantum information technology. Advances in Optics and Photonics 2023, 15 (3), 613−738.

(11) Wang, H.; Qin, J.; Ding, X.; Chen, M.-C.; Chen, S.; You, X.; He, Y.-M.; Jiang, X.; You, L.; Wang, Z. Boson sampling with 20 input photons and a 60-mode interferometer in a 1 0 14-dimensional hilbert space. Physical review letters 2019, 123 (25), 250503.

(12) Senellart, P.; Solomon, G.; White, A. High-performance semiconductor quantum-dot single-photon sources. Nat. Nanotechnol 2017, 12 (11), 1026−1039.

(13) Uppu, R.; Pedersen, F. T.; Wang, Y.; Olesen, C. T.; Papon, C.; Zhou, X.; Midolo, L.; Scholz, S.; Wieck, A. D.; Ludwig, A.; Lodahl, P. Scalable integrated single-photon source. Science advances 2020, 6 (50), No. eabc8268.

(14) Somaschi, N.; Giesz, V.; De Santis, L.; Loredo, J. C.; Almeida, M. P.; Hornecker, G.; Portalupi, S. L.; Grange, T.; Antón, C.; Demory, J.; Gómez, C.; Sagnes, I.; Lanzillotti-Kimura, N. D.; Lemaítre, A.; Auffeves, A.; White, A. G.; Lanco, L.; Senellart, P. Near-optimal single-photon sources in the solid state. Nature Photonics 2016, 10, 340-345

(15) Gazzano, O.; Michaelis de Vasconcellos, S.; Arnold, C.; Nowak, A.; Galopin, E.; Sagnes, I.; Lanco, L.; Lemaître, A.; Senellart, P. Bright solid-state sources of indistinguishable single photons. Nature Communications 2013, 4, 1425

(16) Shirane, M.;Igarashi, Y.; Ota, Y.;Nomura, M.; Kumagai, N.; Ohkouchi, S.; Kirihara, A.;Ishida, S.;Iwamoto, S.;Yorozu, S.; Arakawa, Y. Charged and neutral biexciton–exciton cascade in a single quantum dot within a photonic bandgap. Physica E: Low-dimensional Systems and Nanostructures 2010, 42, 2563-2566.

(17) Kettler, J.; Paul, M.; Olbrich, F.; Zeuner, K.;Jetter, M.; Michler, P.; Florian, M.;Carmesin, C.; Jahnke, F. Neutral and charged biexciton-exciton cascade in near-telecom-wavelength quantum dots. Physical Review B 2016, 94 (4), 045303

(18) Zhai, L.; Löbl, M. C.; Nguyen, G. N.; Ritzmann, J.; Javadi, A.; Spinnler, C.; Wieck, A. D.; Ludwig, A.; Warburton, R. J. Low-noise GaAs quantum dots for quantum photonics. Nature Communications. 2020, 11 (1), 4745

(19) Prescott, S.; Iyer, P. P.; Park, S.; Malek, S.; Noh, J.; Chen, P.; Doiron, C. F.; Addamane, S; Brener, I; Mitrofanov, O. Voltage-Tunable Nonlocal Metasurface for Enhanced Outcoupling of Emission from Quantum Dots. Nano Letters 2026. Accepted

(20) Iyer, P. P.; DeCrescent, R. A.; Mohtashami, Y.; Lheureux, G.; Butakov, N. A.; Alhassan, A.; Weisbuch, C.; Nakamura, S.; DenBaars, S. P.; Schuller, J. A. Unidirectional luminescence from InGaN/GaN quantum-well metasurfaces. Nature Photonics 2020, 14 (9), 543-548.

(21) Curto, A. G.; Volpe, G.; Taminiau, T. H.; Kreuzer, M. P.; Quidant, R.; van Hulst, N. F. Unidirectional Emission of a Quantum Dot Coupled to a Nanoantenna, Science 2010, 329 (5994), 930-933.

(22) Paniagua-Domínguez, R.; Yu, Y. F.; Khaidarov, E.; Choi, S.; Leong, V.; Bakker, R. M.; Liang, X.; Fu, Y. H.; Valuckas, V.; Krivitsky, L. A.; Kuznetsov, A. I. A metalens with a near-unity numerical aperture. Nano Lett. 2018, 18, 2124–2132.

(23) Fang, H. H.; Han, B.; Robert, C.; Semina, M. A.; Lagarde, D.; Courtade, E.; Taniguchi, T.; Watanabe, K.; Amand, T.; Urbaszek, B.; Glazov, M. M.; Marie, X. Control of the Exciton Radiative Lifetime in van der Waals Heterostructures. Phys. Rev. Lett. 2019, 123, 067401.

(24) Bucher, Tobias; Vaskin, A.;Mupparapu, R.; Löchner, F. J. F.;George, A.; Chong, K. E.; Fasold, S.; Neumann, C.; Choi, D.-Y.; Eilenberger, F.; Setzpfandt, F.;Kivshar, Y. S.; Pertsch, T.; Turchanin, A.; Staude, I. Tailoring Photoluminescence from MoS2 Monolayers by Mie-Resonant Metasurfaces. ACS Photonics 2019, 6 (4), 1002-1009

(25) Park, S.; Kim, D.; Seo, M.-K.. Plasmonic Photonic Crystal Mirror for Long-Lived Interlayer Exciton Generation. ACS Photonics 2019, 8 (12), 3619-3626.

(26) Park, S.; Kim, D.; Choi, Y.-S.; Baucour, A.; Kim, D.; Yoon, S.; Watanabe, K.; Taniguchi, T.; Shin, J.; Kim, J.; Seo, M.-K. Customizing Radiative Decay Dynamics of Two-Dimensional Excitons via Position- and Polarization-Dependent Vacuum-Field Interference. Nano Letters 2023, 23 (6), 2158-2165

(27) Park, S.; Kim, J.; Kim, D.; Watanabe, K.; Taniguchi, T.; Seo, M.-K. Demonstration of Two-Dimensional Exciton Complex Palette. ACS Nano 2024, 18 (7), 5647-5655.

(28) Prescott, S.; Iyer, P. P.; Addamane, S.; Jung, H.; Luk, T. S.; Brener, I.; Mitrofanov, O. Mie metasurfaces for enhancing photon outcoupling from single embedded quantum emitters. Nanophotonics 2025, 14 (11), 1917-1925.

(29) Iyer, P. P.; Prescott, S.; Addamane, S.; Jung, H.; Renteria, E.; Henshaw, J.; Mounce, A.; Luk, T. S.; Mitrofanov, O.; Brener, I. Control of Quantized Spontaneous Emission from Single GaAs Quantum Dots Embedded in Huygens' Metasurfaces. Nano Letters 2024, 24 (16), 4749-4757.

(30) Murai, S.; Castellanos, G. W.; Raziman, T. V.; Curto, A. G.; Rivas, J. G. Enhanced Light Emission by Magnetic and Electric Resonances in Dielectric Metasurfaces. Advanced Optical Materials 2020, 8 (16), 1902024.

(31) Capretti, A.; Lesage, A.; Gregorkiewicz, T. Integrating Quantum Dots and Dielectric Mie Resonators: A Hierarchical Metamaterial Inheriting the Best of Both. ACS Photonics 2017, 4 (9) 2187-2196.

(32) Yuan, Y.; Qian, C.; Yang, L.; Ru, X.-C.; Li, Y.; Yang, J.; Fu, B.; Yan, S.; Li, H.; Zuo, Z.; Wang, C.; Hu, X.; Yao, H.-B.; Jin, K.; Gong, Q.; Xu, X. Robust Purcell Effect of CsPbI3 Quantum Dots Using Nonlocal Plasmonic Metasurfaces. Physical Review Letters 2025, 134 (24) 243804.

(33) Park, Y.; Kim, H.; Lee, J.-Y.; Ko, W.; Bae, K.; Cho, K.-S. Direction control of colloidal quantum dot emission using dielectric metasurfaces. Nanophotonics 2020, 9 (5), 1023-1030

(34) Huber, D.; Lehner, B. U.; Csontosová, D.; Reindl, M.; Schuler, S.; Covre da Silva, S. F.; Klenovský, P.; Rastelli, A. Single-particle-picture breakdown in laterally weakly confining GaAs quantum dots. Physical Review B 2019, 100 (23), 235425.

(35) Huo, Y. H.; Witek, B. J.; Kumar, S.; Cardenas, J. R.; Zhang, J. X.; Akopian, N.; Singh, R.; Zallo, E.; Grifone, R.; Kriegner, D.; Trotta, R.; Ding, F.; Stangl, J.; Zwiller, V.; Bester, G.; Rastelli, A.; Schmidt, O. G. A light-hole exciton in a quantum dot. Nature Physics 2014, 10 (1), 46–51.

(36) Abbarchi, M.; Kuroda, T.; Mano, T.; Sakoda, K.; Mastrandrea, C. A.; Vinattieri, A.; Gurioli, M.; Tsuchiya, T. Energy renormalization of exciton complexes in GaAs quantum dots. Physical Review B 2010, 82 (20), 201301.

(37) Fu, Y. J.; Lin, S. D.; Tsai, M. F.; Lin, H.; Lin, C. H.; Chou, H. Y.; Cheng, S. J.; Chang, W. H. Anomalous diamagnetic shift for negative trions in single semiconductor quantum dots. Physical Review B 2010, 81 (11), 113307.

(38) Decker, M.; Staude, I.; Falkner, M.; Dominguez, J.; Neshev, D. N.; Brener, I.; Pertsch, T.; Kivshar, Y. S. High-Efficiency Dielectric Huygens' Surfaces. Advanced Optical Materials 2015, 3 (6), 813-820.

(39) Malek, S. C.; Norden, T.; Doiron, C. F.; Santiago-Cruz, T.; Yu, J.; Cerjan, A.; Padmanabhan, P.; Brener, I. Giant Enhancement of Four-Wave Mixing by Doubly Zone-Folded Nonlocal Metasurfaces. ACS Nano 2025, 19 (40), 35609-35617.

(40) Malek, S. C.; Doiron, C. F.; Brener, I.; Cerjan, A. Robust multiresonant nonlocal metasurfaces by rational design. Nanophotonics 2025, 14 (4), 449-458.

(41) Abbarchi, M.; Troiani, F.; Mastrandrea, C.; Goldoni, G.; Kuroda, T.; Mano, T.; Sakoda, K.; Koguchi, N.; Sanguinetti, S.; Vinattieri, A.; Gurioli, M. Spectral diffusion and line broadening in single self-assembled GaAs/AlGaAs quantum dot photoluminescence. Applied physics letters 2008, 93, 162101

(42) Basso Basset, F.; Bietti, S.; Tuktamyshev, A.; Vichi, S.; Bonera, E.; Sanguinetti, S. Spectral broadening in self-assembled GaAs quantum dots with narrow size distribution. J. Appl. Phys. 2019, 126, 024301

(43) Rakhlin, M.; Belyaev, K.; Klimko, G.; Mukhin, I.; Kirilenko, D.; Shubina, T.; Ivanov, S.; Toropov, A. InAs/AlGaAs quantum dots for single-photon emission in a red spectral range. Sci. Rep. 2018, 8 (1), 5299.

(44) Zhou, T.; Tang, M.; Xiang, G.; Xiang, B.; Hark, S.; Martin, M.; Baron, T.; Pan, S.; Park, J.-S.; Liu, Z. Continuous-wave quantum dot photonic crystal lasers grown on on-axis Si (001). Nat. Commun. 2020, 11 (1), 977.

(45) Wang, Z.; Rasmita, A.; Long, G.; Chen, D.; Zhang, C.; Garcia, O. G.; Cai, H.; Xiong, Q.; Gao, W. Optically driven giant superbunching from a single perovskite quantum dot. Advanced Optical Materials 2021, 9 (21), 2100879.

(46) Meuret, S.; Tizei, L. H. G.; Cazimajou, T.; Bourrellier, R.; Chang, H. C.; Treussart, F.; Kociak, M. Photon Bunching in Cathodoluminescence. Physical Review Letters 2015, 114 (19), 197401.

(47) Kuroda, T.; Belhadj, T.; Abbarchi, M.; Mastrandrea, C.; Gurioli, M.; Mano, T.; Ikeda, N.; Sugimoto, Y.; Asakawa, K.; Koguchi, N.; Sakoda, K.; Urbaszek, B.; Amand, T.; Marie, X. Bunching visibility for correlated photons from single GaAs quantum dots. Physical Review B 2009, 79 (3), 035330.

(48) Park, S.; Azizur-Rahman, K. M.; Shima, D.; Balakrishnan, G.; Yu, J.; Jung, H.; Mah, J. J.; Prescott, S.; Chen, P.; Addamane, S.; Pete, D.; Mounce, A.; Luk, T. S.; Iyer, P. P.; Brener, I.; Mitrofanov, O. Efficient single-photon emission via quantum-confined charge funneling to quantum dots. Communication Materials 2025, 6 (1), 286.

(49) Jin. W.; Li. W.; Orenstein. M.; Fan. S. Inverse design of lightweight broadband reflector for relativistic lightsail propulsion. ACS Photonics 2020, 7 (9), 2350-2355.

(50) Zanotti, S.; Minkov, M.; Nigro, D.; Gerace, D.; Fan, S.; Andreani, L. C. Legume: A free implementation of the guided-mode expansion method for photonic crystal slabs. Computer Physics Communications 2024, 304, 109286.

(51) Aspnes, D. E.; Kelso, S. M.; Logan, R. A.; Bhat, R. Optical properties of $Al_xGa_{1-x}$ As. Journal of Applied Physics 1986, 60 (2), 754-767.

# Supplementary Information

# Simultaneous anti-bunched and super-bunched photons from a GaAs Quantum dot in a dielectric metasurface

**Sanghyeok Park,[1,2] Oleg Mitrofanov,[3] Kusal M. Abeywickrama,[4] Samuel Prescott,[3] Jaeyeon Yu,[1,2] Stephanie C Malek,[1,2] Hyunseung Jung,[1,2] Emma Renteria,[5] Sadhvikas Addamane,[1,2] Alisa Javadi,[4] Igal Brener,[1,2,*] and Prasad P Iyer[1,2,*]**

[1]*Sandia National Laboratories, Albuquerque, New Mexico 87185, United States*
[2]*Center for Integrated Nanotechnologies, Sandia National Lab, Albuquerque, New Mexico 87185, United States*
[3]*University College London, Electronic and Electrical Engineering, London WC1E 7JE, U.K*
[4]*Department of Physics and Astronomy, Center for Quantum Research and Technology, University of Oklahoma, Norman, Oklahoma 73019, United States*
[5]*Center for High Technology Materials, University of New Mexico, Albuquerque, New Mexico 87185, United States*
*[\*ibrener@sandia.gov](mailto:ibrener@sandia.gov), [ppadma@sandia.gov](mailto:ppadma@sandia.gov)*

## S1. Enhancement of PL from exciton complexes in GaAs quantum dot embedded in metasurface

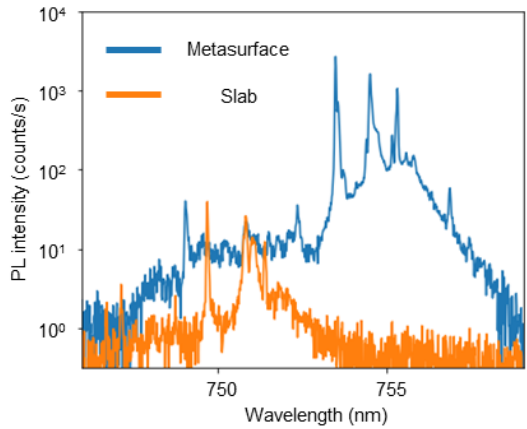

Figure S1. Photoluminescence spectra comparison between a quantum dot embedded in a dielectric Mie metasurface (blue) and in an unpatterned slab (orange).

Figure S1 compares the photoluminescence (PL) spectra of quantum dots (QDs) embedded in a dielectric Mie metasurface versus those in a bare slab. To enable direct comparison, the PL intensities are normalized to the pump power. While both samples display similar spectral features corresponding to excitonic transitions, the metasurface-embedded QDs exhibit over an order of magnitude higher PL intensities across the full spectral range.

## S2. Pump power dependence of QD PL spectra

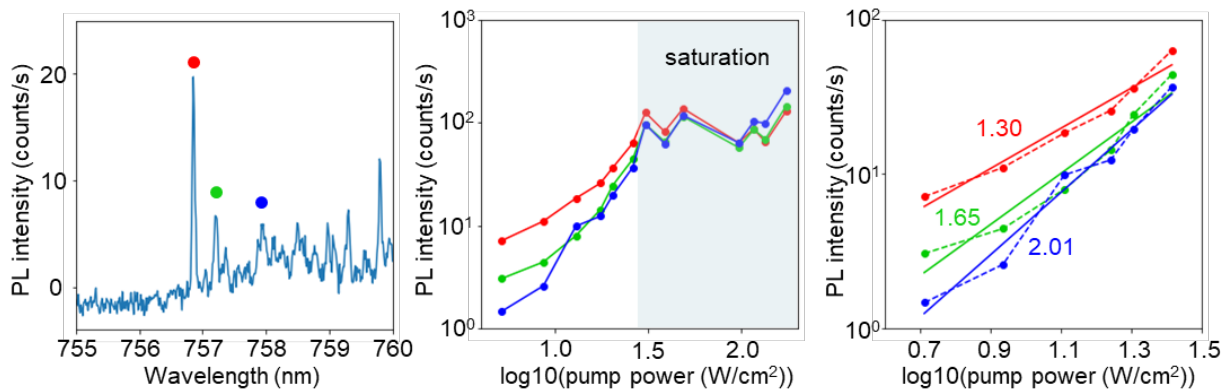

Figure S2. Photoluminescence properties of three positively charged excitonic emission lines from an embedded GaAs QD in a dielectric metasurface. (a) PL spectrum highlighting three excitonic transitions labeled with red, green, and blue markers. (b) Power-dependent PL intensities of the three labeled peaks

plotted on a log-log scale, showing a gradual saturation above ~50 W/cm² (shaded region). (c) Log-log fit of PL intensity versus pump power for the three emission lines below the saturation power. The extracted slopes are 1.30 (red), 1.65 (green), and 2.01 (blue).

We investigated the excitation power dependence of multiple excitonic transitions from a single QD. Figure S2a presents the PL spectrum measured under non-resonant excitation with 8.6 W/cm², with three representative peaks marked in red, green, and blue. These peaks correspond to different positively charged exciton complexes. In Figure S2b, the PL intensities of the three peaks monotonically increase followed by saturation at ~50 W/cm². In the pre-saturation region, the blue peaks exhibit the quadratic pump power dependence. This trend confirms that the blue peak is coming from the charged biexciton [1-4].

## S3. Super-bunching emission from positively charged exciton complexes

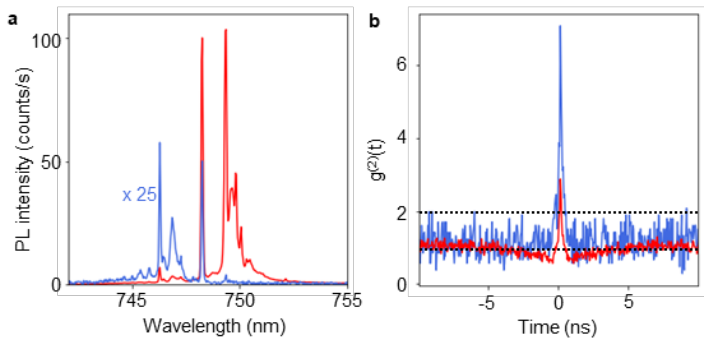

Figure S3. PL spectra and second-order photon correlations from the GaAs QD in Figure 2. (a) Photoluminescence spectra collected from the GaAs QD in the metasurface with different spectral filtering. The PL spectrum without spectral filtering is depicted with the red color and the short pass filtered at 748 nm is depicted with the blue color. For the visibility, the intensity of the short passed one is multiplied by 25. (b) Second-order correlation functions $g^{(2)}(\tau)$ measured for both unfiltered (red) and short pass filtered (blue) cases. A pronounced photon bunching is observed in the short pass filtered one, indicating correlated multi-photon emission.

The QD used in the magneto-PL study also exhibits similar photoluminescence and photon correlation features as those discussed in the main manuscript. In Fig. S3a, the PL spectra with short-pass filtering shows only the emission features at shorter wavelengths than the neutral exciton, primarily associated with positively charged exciton complexes, while the red spectrum includes all excitonic transitions. These short-wavelength features exhibit extremely strong temporal photon correlations, with $g^{(2)}(0)$ significantly exceeding 2, indicating super-bunching (Fig S3b). In contrast, the unfiltered spectrum shows diluted correlation due to averaging across multiple excitonic states just like the second order correlation measurement result depicted with the green dots in Figure 1c. This consistency confirms that the observed photon statistics are not specific to a single dot but representative of our QD-metasurface platform.

## S4. Super-bunching emission from a GaAs QD embedded in Mie metasurface

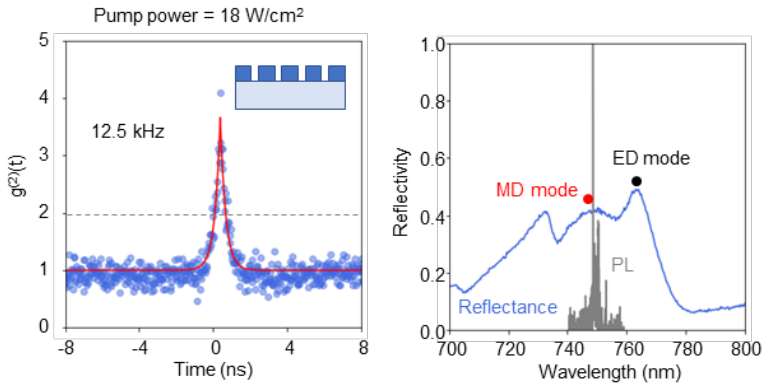

Figure S4. (a) Second-order photon correlation $g^{(2)}(\tau)$ of emission from a quantum dot embedded in a Mie metasurface, showing a pronounced super-bunching peak at zero delay. The count rate is 11.8 kHz. The red line represents a fit based on convolution with the system response. Inset: schematic of the metasurface geometry. (b) Photoluminescence (PL) spectrum (black) and reflectance spectrum (blue) from the same Mie metasurface device, revealing the spectrally broad Mie resonance that enhances multiple excitonic emission peaks.

Figure S4(a) presents the measured $g^{(2)}(\tau)$ function from a QD coupled to a dielectric Mie metasurface. A clear photon super-bunching behavior is observed, with $g^{(2)}(0) = 3.55$, indicating temporally correlated photon bursts. The enhancement of such nonlinear photon statistics is facilitated by the metasurface, which significantly increases emission intensity from weak excitonic complexes.

Figure S4(b) compares the PL spectrum of the QD with the reflectance of the metasurface. The overlapping spectral features suggest that the broad electric and magnetic dipole Mie resonances enhance the emission across multiple excitonic transitions, enabling the observation of complex photon statistics at detectable count rates. The detail mode analysis and the effect of the optical structures are in [5,6].

## S5. Second order correlation function dependence on polarization for QDs in Huygens' metasurface

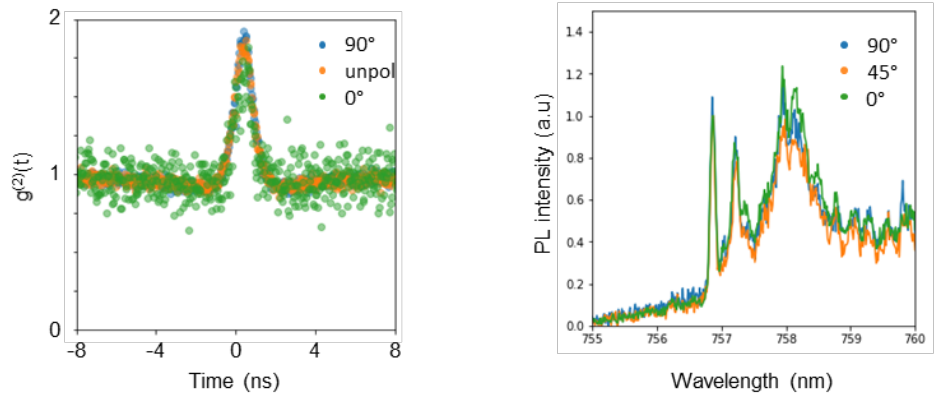

Figure S5. Polarization dependence of second-order correlation and photoluminescence spectra.(a) Second-order correlation functions g(2)(τ) of a selected emission peak measured through a polarizer set at 0°, 90°, and without polarization filtering. All traces exhibit comparable bunching peaks near zero delay, indicating the absence of a preferred linear polarization. (b) Polarization-resolved PL spectra collected through analyzer angles of 0°, 45°, and 90°. The emission intensity and spectral features remain nearly unchanged across polarizations, corroborating the unpolarized nature of the super-bunched photon emission.

Figure S5 presents polarization-resolved measurements to investigate the polarization nature of the super-bunched emission from a QD embedded in a metasurface. The second-order correlation functions $g^{(2)}(\tau)$ (Fig S5a) show nearly identical temporal bunching features regardless of whether the emission is analyzed through 0°, 90°, or no polarizer, suggesting that the underlying emission is not linearly polarized. Similarly, the PL spectra (Fig S5b) taken through different polarization analyzers show consistent spectral shapes and intensity distributions, further supporting the conclusion that the super-bunched photon streams are unpolarized. These results motivate future studies with circular polarization dependent cross-correlation measurements.

# References

[1] Heindel, T., Thoma, A., von Helversen, M. *et al.* A bright triggered twin-photon source in the solid state. *Nat Commun* **8**, 14870 (2017). https://doi.org/10.1038/ncomms14870

[2] Benson, O., Santori, C., Pelton, M. & Yamamoto, Y. Regulated and entangled photons from a single quantum dot. *Phys. Rev. Lett.* **84**, 2513 (2000).

[3] Huber, D., Reindl, M., Huo, Y. *et al.* Highly indistinguishable and strongly entangled photons from symmetric GaAs quantum dots. *Nat Commun* **8**, 15506 (2017). https://doi.org/10.1038/ncomms15506

[4] He, YM., Iff, O., Lundt, N. et al. Cascaded emission of single photons from the biexciton in monolayered $WSe_2$. Nat Commun 7, 13409 (2016). https://doi.org/10.1038/ncomms13409

[5] Prescott, S.; Iyer, P. P.; Addamane, S.; Jung, H.; Luk, T. S.; Brener, I.; Mitrofanov, O. Mie metasurfaces for enhancing photon outcoupling from single embedded quantum emitters. Nanophotonics 2025, 14 (11), 1917-1925.

[6] Iyer, P. P.; Prescott, S.; Addamane, S.; Jung, H.; Renteria, E.; Henshaw, J.; Mounce, A.; Luk, T. S.; Mitrofanov, O.; Brener, I. Control of Quantized Spontaneous Emission from Single GaAs Quantum Dots Embedded in Huygens' Metasurfaces. Nano Letters 2024, 24 (16), 4749-4757.